\documentclass[10pt,conference]{IEEEtran}
\IEEEoverridecommandlockouts
\usepackage{cite}
\usepackage{amsmath,amssymb,amsfonts}
\usepackage{algorithmic}
\usepackage{graphicx}
\usepackage{textcomp}
\usepackage{xcolor}
\usepackage{balance}
\usepackage{booktabs}
\usepackage{xspace}
\usepackage{listings}
\usepackage[frozencache,cachedir=_minted]{minted}
\usepackage{seqsplit}
\usepackage[hidelinks]{hyperref}
\usepackage{xparse}
\usepackage{enumitem}
\usepackage{tikz}
\usetikzlibrary{arrows.meta,positioning,fit,backgrounds,calc}
\usepackage{comment} 
\usepackage{inconsolata}
\usepackage{xurl}
\usepackage{subcaption}
\newcommand{\sysname}{\textsc{WorkflowBP}\xspace}

\newcounter{finding}
\newcommand{\finding}[1]{%
  \par\smallskip\smallskip\noindent
  {\centering
   {\setlength{\fboxsep}{2pt}%
    \fbox{\setlength{\fboxsep}{4pt}%
      \fbox{\parbox{0.95\columnwidth}{%
        \refstepcounter{finding}%
        \textbf{Finding \thefinding.} #1}}}}\par\smallskip}%
}

\newcommand{\etal}{et al.}
\newcommand{\eg}{e.g.}
\newcommand{\ie}{i.e.}

\newcommand{\ccTitle}[0]{Missing Security-Fix Backporting for \\ GitHub Actions Workflows}

\setminted[yaml]{
    frame=lines,
    framesep=2mm,
    baselinestretch=1.2,
    xleftmargin=0.5em,
    bgcolor=gray!5,
    fontsize=\scriptsize,
    linenos=true,     
    numbersep=4pt,
    tabsize=2,
    breaklines=true,         
    breakanywhere=true,      
    breaksymbol=\color{gray}\ensuremath{\hookrightarrow} 
}

\setminted[diff]{
    frame=lines,
    framesep=2mm,
    baselinestretch=1.2,
    xleftmargin=0.5em,
    bgcolor=gray!5,
    fontsize=\scriptsize,
    linenos=true,
    numbersep=4pt,
    tabsize=2,
    breaklines=true,
    breakanywhere=true,
    breaksymbol=\color{gray}\ensuremath{\hookrightarrow}
}

\setminted[text]{
    frame=lines,
    framesep=2mm,
    baselinestretch=1.2,
    xleftmargin=0.5em,
    bgcolor=gray!5,
    fontsize=\scriptsize,
    linenos=false,
    numbersep=4pt,
    tabsize=2,
    breaklines=true,
    breakanywhere=true,
    breaksymbol=\color{gray}\ensuremath{\hookrightarrow}
}

\def\BibTeX{{\rm B\kern-.05em{\sc i\kern-.025em b}\kern-.08em
    T\kern-.1667em\lower.7ex\hbox{E}\kern-.125emX}}
\begin{document}

\title{\ccTitle}

\makeatletter
\newcommand{\linebreakand}{%
  \end{@IEEEauthorhalign}
  \hfill\mbox{}\par
  \mbox{}\hfill\begin{@IEEEauthorhalign}
}
\makeatother

\author{\IEEEauthorblockN{Chengyan Ma\IEEEauthorrefmark{1}, Jieke Shi\IEEEauthorrefmark{1}\thanks{\textit{Corresponding author: Jieke Shi.}}, 
Ruidong Han\IEEEauthorrefmark{1}, Weifeng Sun\IEEEauthorrefmark{1},
Yuqing Niu\IEEEauthorrefmark{1}, Yunbo Lyu\IEEEauthorrefmark{1},
Zhou Yang\IEEEauthorrefmark{2}\IEEEauthorrefmark{3}\IEEEauthorrefmark{4}, David Lo\IEEEauthorrefmark{1}}
\IEEEauthorblockA{\IEEEauthorrefmark{1}\textit{Singapore Management University}, Singapore~
\IEEEauthorrefmark{2}\textit{University of Alberta}, Edmonton, Canada\\
\IEEEauthorrefmark{3}\textit{Alberta Machine Intelligence Institute}, Edmonton, Canada~
\IEEEauthorrefmark{4}\textit{Canada CIFAR AI Chair}, Canada\\
\{chengyanma, jiekeshi, rdhan, wfsun, yuqingniu, yunbolyu\}@smu.edu.sg,
zy25@ualberta.ca, davidlo@smu.edu.sg}
}

\maketitle

\begin{abstract}
GitHub natively supports workflow automation through GitHub Actions, now the dominant CI/CD tool on the platform. Yet keeping these workflows secure remains an ongoing burden: maintainers often patch a vulnerability on the default branch but leave the same flaw on the release and maintenance branches that downstream consumers continue to build from. Carrying a known fix onto these divergent branches, called \emph{backporting}, is rarely done, and this paper presents the first large-scale empirical study quantifying the resulting gap. From a public corpus of GitHub Actions workflow histories, we mine 1,903,191 workflow-modifying commits, scan each before and after with the static scanner \textsc{zizmor} to identify 69,865 security fixes across 24,238 projects, and audit every release branch to check whether each fix was carried over. The gap is pervasive: 224,392 of 474,490 audited release branches (47.3\%) remain vulnerable, and deliberate backporting is rare and slow, with only 19,476 of 78,907 already-fixed branches confirmed as plausible backports at a median lag of 243 days. We thus design \sysname, the first technique for automated security-fix backporting of GitHub Actions workflows. \sysname takes a neuro-symbolic approach: it compiles a default-branch fix into a domain-specific representation called the Workflow Semantic Patch (WSP) and applies it to each divergent branch by matching jobs and steps based on their content rather than their position, while an LLM synthesizes the WSP for the most divergent cases the deterministic engine cannot resolve. A backport is accepted only after re-scanning confirms the targeted vulnerability is fixed and nothing else broke. On 12,000 real unpatched workflow files from our study, \sysname produces a scanner-checked backport for 67.9\%, far outperforming verbatim copy-paste (18.0\%), Dependabot-style updates (13.0\%), and an LLM-only baseline (49.5\%). We further submit 176 issues with \sysname-generated backports to real-world projects, 44 of which have been confirmed or fixed by developers.
\end{abstract}


\section{Introduction}\label{sec:intro}

In November 2019, GitHub released GitHub Actions\footnote{\url{https://github.com/features/actions}}, a cloud-native tool that integrates Continuous Integration, Deployment, and Delivery (CI/CD) into its platform, letting developers write a workflow as a YAML file that automates tasks in response to repository events, for instance, running the test suite on every pull request or building and publishing a release artifact on a version tag. It has since become the dominant workflow automation tool on GitHub, supporting more than 3M workflows across over 49.2K public repositories~\cite{10.1145/3643991.3644867,9825792,9978190}. This convenience, however, comes with security risk~\cite{wiz-tjactions,cycode-tjactions}: a workflow that builds a release or deployment can silently pull in a compromised third-party component and ship an artifact that carries the flaw. This is also exemplified by several documented vulnerabilities such as CVE-2025-30066~\cite{cve-2025-30066}, where a threat actor rewrote the release tags of the widely used \texttt{\small{tj-actions/changed-files}} action to point at a commit containing malicious code~\cite{wiz-tjactions}, so that every workflow referencing the action by tag executed that code on its next run, leaking CI/CD secrets across thousands of repositories.

\begin{figure}[!t]
\centering
\begin{minted}[fontsize=\scriptsize,frame=none,escapeinside=||,xleftmargin=0.5em]{yaml}
# (a) main, before: vulnerable!
- uses: actions/checkout@v4        # mutable tag vulnerability
\end{minted}
\begin{minted}[fontsize=\scriptsize,frame=none,escapeinside=||,xleftmargin=0.5em]{yaml}
# (b) main, after: hardened
- uses: actions/checkout@8ade135   # pinned to an immutable commit ID
  with:
    persist-credentials: false
\end{minted}
\begin{minted}[fontsize=\scriptsize,frame=none,escapeinside=||,xleftmargin=0.5em]{yaml}
# (c) release/v1.x: still vulnerable (never backported)!
- uses: actions/checkout@v4        # same mutable tag vulnerability
\end{minted}
\caption{The backporting gap in miniature. A maintainer hardens
\texttt{actions/checkout} on the default branch (a\,$\rightarrow$\,b) by pinning
the mutable tag to an immutable commit ID, yet this fix is never backported to
the \texttt{release/v1.x} branch (c), which still references \texttt{@v4} and
remains exposed to the tag-rewrite attack of CVE-2025-30066~\cite{cve-2025-30066}.}
\label{fig:motivating}
\end{figure}

Such a flaw is of course fixable, but a factor that is easy to overlook is how open-source projects are maintained. A project rarely lives on a single branch~\cite{8854734,bunyakiati2020opensourceprojectscollaborativedevelopment}: alongside the default branch, maintainers keep long-lived release and maintenance branches so that each shipped version can receive fixes without being forced onto the latest code, and downstream users pin to and build from those branches for stability~\cite{10.1145/3460319.3464821,8816809}. Fixing only the default branch is therefore insufficient, as the vulnerable workflow persists on every release branch that was not touched. Indeed, the security analyses following the \texttt{\small tj-actions} incident stressed that patching the default branch alone leaves exposed the older branches and released versions that downstream users still build from~\cite{wiz-tjactions,cycode-tjactions}. Fig.~\ref{fig:motivating} shows this concretely: the maintainer repins \texttt{\small{actions/checkout}} to an immutable commit ID on the default branch, but the \texttt{\small{release/v1.x}} branch still references the mutable tag \texttt{\small @v4} and remains exposed.

Carrying a fix onto the branches that remain exposed is called \emph{backporting}, i.e., transferring a security fix that already exists on one branch onto the divergent branches that still carry the weakness. Prior studies~\cite{10.1145/2771783.2771796,10.1145/3551349.3556893,10.1145/3412376,10.1145/3460319.3464821,10.1145/3576915.3623188} of backporting in other settings, most notably Linux kernel security patches~\cite{10.1145/3460319.3464821,10.1145/3576915.3623188}, have found that developers frequently fail to propagate fixes to all the branches that need them. Whether the same gap afflicts GitHub Actions workflows, however, is unknown, and it is the question this paper sets out to answer.

Our study addresses this open question with the first investigation of  how often security fixes fail to propagate across branches in open-source projects that use GitHub Actions. From the largest public corpus of workflow histories~\cite{10.1145/3643991.3644867}, collected between 2024 and 2026, we retain the 1{,}903{,}191 workflow-modifying commits across 24{,}238 projects, and scan each one before and after the change with \textsc{zizmor}~\cite{zizmor}, a static analyzer for GitHub Actions. We label a commit a \emph{security fix} when it removes a weakness and introduces none, yielding 69{,}865 fixes. Since these fixes land on the default branch, they protect only the workflows built from that branch, so we then ask whether they also reach the release and maintenance branches that downstream users continue to build from. We re-scan each such branch and, where the weakness is gone, compare when it disappeared against the timing of the default-branch fix, which lets us identify a deliberate backport, i.e., a later and separate remediation. Our results show that the gap is wide and slow to close: of 474{,}490 release branches, 224{,}392 (47.3\%) remain vulnerable, and of those eventually fixed, only 19{,}476 received a deliberate backport, arriving at a median lag of 243 days.

A gap of this scale cannot realistically be closed by hand, which raises the question of whether workflow backporting can be automated. Existing backporting techniques, however, do not transfer to GitHub Actions. Prior work on transplantation and backporting~\cite{6606596,10.1145/2771783.2771796,10.1145/3551349.3556893,10.1145/3412376,10.1145/3460319.3464821,10.1145/3576915.3623188} mostly builds on programming-language syntax matching or even semantics, such as reasoning over and editing abstract syntax trees, driven by type and data-flow reasoning and validated against a compiler and test suite, none of which a YAML workflow, being a declarative configuration format rather than a general-purpose language, exposes. Dependency updaters such as Dependabot~\cite{dependabot} and Renovate~\cite{renovate} are the only backport-like tools, yet they move a single dependency forward to a newer version and can neither couple edits across multiple locations in a workflow file nor move a fix backward onto an older branch. We therefore run the applicable options on our datasets: a heuristic copy-paste of the source diff, a Dependabot-style update, and a large language model, which has shown exceptional capability in programming~\cite{10.1145/3695988}, prompted to rewrite the file. Our empirical evaluation confirms the gap: these methods repair only 18.0\%, 13.0\%, and 49.5\% of files respectively (\S\ref{sec:eval}), still leaving most branches unbackported.

Reading these failures exposes three challenges that any workflow backporting tool must overcome. \emph{Multiplicity}: one fix is missing on a mean of 31.3 release branches per project (up to 2,971), so it must be replayed across many branches. \emph{Coupling}: 34.3\% of clean fixes bundle two or more weakness types into one commit whose edits are correct only together. \emph{Structural divergence}: on over 60\% of unpatched branches, the fix's original location no longer exists, as release branches rename jobs, reorder steps, and refactor files. Each baseline stumbles on one of these, i.e., divergence defeats copy-paste, coupling defeats the updater, and unresolved external state defeats the LLM, and no tool addresses all three, which is the gap \sysname fills.

\smallskip
\noindent\textbf{\sysname.} We present \sysname, the first neuro-symbolic technique for automated security-fix backporting of GitHub Actions workflows. Given a security fix already applied on the default branch, \sysname \emph{compiles} the fixing commit into a \emph{Workflow Semantic Patch} (WSP), a domain-specific representation that records the fix as a compact program of add, delete, and rewrite edits over the workflow's structure rather than its text; \emph{applies} this single WSP to each vulnerable release branch by binding its anchors to whatever the branch actually contains; and \emph{validates} the result, accepting a patch only when four independent oracles agree that the targeted weakness is gone and nothing else broke. For the most divergent fixes that the deterministic engine cannot place, \sysname falls back to a large language model under counterexample-guided inductive synthesis (CEGIS), which proposes a revised WSP from the engine's counterexamples but never edits the file itself, so every accepted patch is validated by re-scanning rather than trusted. 

Our WSP thus resolves all three challenges at once: it applies as a single atomic unit, so a coupled fix never lands half-applied (Coupling); its edits anchor to each job and step by \emph{content}, i.e., the job name and the action it runs, rather than by position, so it still resolves when jobs are renamed and steps reordered (Structural divergence); and it fans out autonomously across a project's hundreds of affected branches, each validated on its own (Multiplicity). 
On 12,000 real unpatched files from our study, \sysname produces a scanner-checked backport for 67.9\%, and 44 of 176 submitted as issues were confirmed or fixed by maintainers.

The main contributions of this paper are:
\begin{itemize}[leftmargin=*]
\item \textbf{Study.} The first large-scale study of the security-fix backporting gap in GitHub Actions workflows (24,238 projects, 474,490 branches), showing it pervasive (47.3\% unpatched) and slow (median 243 days).
\item \textbf{Benchmark.} A reusable dataset of security fixes, unpatched branches, and confirmed backports mined from real histories, as ground truth for evaluating backporting techniques.
\item \textbf{Approach.} \sysname, the first neuro-symbolic technique to backport a coupled workflow fix onto a divergent branch as one patch: a deterministic engine compiles the fix into a Workflow Semantic Patch and matches jobs and steps by content, an LLM resolves the most divergent cases, and every patch is validated by re-scanning.
\item \textbf{Evaluation.} We evaluate \sysname on 12,000 real unpatched files against copy-paste, Dependabot-style, and LLM baselines.
\item \textbf{Practice.} We submit 176 backports as issues to real-world projects, 44 confirmed by maintainers.
\end{itemize}

\section{Background and Motivations}
\label{sec:background}

\subsection{GitHub Actions in Brief}
\label{sec:gha-primer}

\begin{figure}[t!]
\centering
\begin{minted}[style=colorful]{yaml}
name: CI
on: [push, pull_request]        # run on every push and PR
jobs:
  build:
    runs-on: ubuntu-latest
    steps:
      - uses: actions/checkout@v4          # check out the source
      - name: Install build dependencies
        run: sudo apt-get install -y libssl-dev
      - name: Compile and run the test suite
        run: ./configure && make && make check
      - uses: codecov/codecov-action@v4    # upload coverage
\end{minted}
\caption{An example GitHub Actions CI workflow, run on every push and pull
request. It composes two third-party actions, \texttt{actions/checkout} and
\texttt{codecov/codecov-action}; each \texttt{uses:} clause is a point at which
an external, trust-dependent component enters the pipeline.}
\label{fig:ga-primer}
\end{figure}

GitHub Actions is a workflow automation feature built into GitHub that lets maintainers define pipelines which run in response to events such as commits and pull requests, automating tasks like testing, building, and deployment~\cite{9463074,9978190}. Workflows are written as YAML files under \texttt{\small .github/workflows/}, and Fig.~\ref{fig:ga-primer} shows one for continuous integration. The \texttt{\small on:} field is the \emph{trigger}, i.e., the events that start a run, so here every \texttt{\small push} and pull request launches the pipeline. A workflow contains one or more \emph{jobs}, each running on a fresh virtual machine selected by \texttt{\small runs-on}, and each job is an ordered list of \emph{steps}. A step either invokes a reusable component called an \emph{action} through a \texttt{\small uses:} clause, optionally configured with \texttt{\small with:} inputs, or runs a shell command through a \texttt{\small run:} clause. The example checks out the repository, installs dependencies, builds and tests the project, and uploads a coverage report, composing the third-party actions \texttt{\small actions/checkout} and \texttt{\small codecov/codecov-action}. This composable syntax has made GitHub Actions the most widely used CI/CD service on GitHub~\cite{10.1145/3643991.3644867,ROSTAMIMAZRAE2026112824,10.1145/3597503.3623303,10.1145/3736758}, relied on by many large and security-critical projects for continuous integration~\cite{10.1145/3749371,281396,moazen2025granitegranularruntimeenforcement}.

\subsection{Workflow Security Weaknesses and the Backporting Gap}
\label{sec:vuln-taxonomy}

The routine workflow in Fig.~\ref{fig:ga-primer} already carries three weaknesses that recur across our corpus. \emph{Unpinned actions} (\texttt{\small unpinned-uses}) are referenced by a mutable tag rather than an immutable SHA, so whoever controls the tag controls what every consumer runs, the mechanism behind CVE-2025-30066~\cite{cve-2025-30066}. \emph{Excessive token permissions} (\texttt{\small excessive-permissions}) arise when a missing \texttt{\small permissions} block lets \texttt{\small GITHUB\_TOKEN} inherit repository-wide write access. \emph{Persisted credentials} (\texttt{\small artipacked}) occur because \texttt{\small actions/checkout} writes the token to disk unless \texttt{\small persist-credentials: false} is set. These weaknesses are seldom fixed one at a time: a single commit usually addresses all three, and the remediation is complete only when every edit lands together, since pinning the actions while leaving credentials persisted still leaks the token. Such a fix is typically applied only to the default branch, while the same vulnerable workflow lingers on the release branches downstream users build from. Carrying it onto those branches, i.e., \emph{backporting}, must reproduce the fix's security intent in each branch's own context rather than copy its text, since release branches diverge in action versions, runtime, and job structure. Divergence also blurs what counts as a backport, as a release branch mixes forward merges, cherry-picks, and branch-specific patches, so a present fix may reflect an incidental auto-merge rather than a deliberate patch, a distinction we make precise in \S\ref{sec:study:rq2}.

\section{The Backporting Gap: Empirical Studies}\label{sec:study}

Although workflow security fixes are often applied to the default branch, little is known about whether these fixes are consistently propagated to the release branches that downstream users continue to rely on.
We therefore conduct a large-scale empirical study of how security fixes in GitHub Actions workflows propagate from the default branch to release branches.
The study is organized around three research questions and produces two ground-truth artifacts, namely unfixed branches and confirmed backports, that we later use to evaluate \sysname.

\begin{itemize}[leftmargin=*]
  \item \textbf{RQ1 (Gap).} Do security fixes on the default branch reach the release branches that downstream users build from?
  \item \textbf{RQ2 (Practice).} How often, and how quickly, are release branches fixed by deliberate backporting?
  \item \textbf{RQ3 (Challenges).} What dimensions of difficulty must an automated backporting tool overcome?
\end{itemize}

\subsection{Data Collection}\label{sec:study:design}

All three research questions build on a common pipeline, described next, that turns raw workflow commits into a labeled set of \textit{clean security fixes}.

\smallskip
\noindent\textbf{Commit selection.} We start from a public dataset of GitHub Actions workflow histories~\cite{10.1145/3643991.3644867} and retain entries that modify workflow files under \texttt{\small .github/workflows/}. This dataset is mined from real-world public repositories, not synthetic or toy projects: it records the full revision history of more than 116,000 workflow files across over 25,000 active public GitHub repositories~\cite{10.1145/3643991.3644867}, so the fixes we study and the release branches we audit come from projects that downstream users actually depend on. The dataset contains 3,418,911 file-level workflow changes across the corpus. We exclude entries that (i) add or delete files, for which paired pre- and post-change workflow contents are unavailable; (ii) correspond to rename or copy operations; or (iii) contain YAML that cannot be parsed or is not recognizable as a GitHub Actions workflow. After filtering, our working set contains 1,903,191 workflow-changing commits, which we use throughout the remainder of \S\ref{sec:study:design}.

\smallskip
\noindent\textbf{Before/after scanning.}
For each commit, we run the static workflow security scanner \textsc{zizmor}~\cite{zizmor} on the workflow file in both its pre-commit and post-commit state, obtaining two sets of security issues present before and after the change.

\smallskip
\noindent\textbf{Security-fix filter.}
Let $B$ and $A$ be the sets of security issues \textsc{zizmor} reports on the workflow before and after the commit, respectively. From them we derive the issues the commit \textit{removes}, $V_\mathit{fixed} = B \setminus A$ (present before, absent after), and the issues it \textit{introduces}, $V_\mathit{introduced} = A \setminus B$ (absent before, present after).
We label a commit a \textit{security fix} when it removes at least one issue and introduces none: (i)~$V_\mathit{fixed} \neq \varnothing$ and (ii)~$V_\mathit{introduced} = \varnothing$. We adopt this two-part criterion for a specific reason. Condition~(i) requires the commit to remove a real weakness, so that we study genuine fixes rather than unrelated edits. Condition~(ii) requires it to introduce no new weakness, so that the change we later backport has an unambiguous security effect, i.e., a pure improvement rather than one that trades one weakness for another.

\subsection{RQ1: How Wide Is the Cross-Branch Backporting Gap?}\label{sec:study:rq1}

\begin{table}[t]
  \centering
  \caption{Outcome of auditing every release branch of the 24,238 projects
  against the corresponding default-branch fix.}
  \label{tab:branches}
  \begin{tabular}{lrr}
    \toprule
    \textbf{Outcome} & \textbf{Branches} & \textbf{Share} \\
    \midrule
    Inapplicable (file absent)      & 171,191 & 36.1\% \\
    Already-fixed                   & 78,907 & 16.6\% \\
    Still vulnerable (\textsc{gap}) & 224,392 & 47.3\% \\
    \midrule
    Total release branches audited  & 474,490 & 100\% \\
    \bottomrule
  \end{tabular}
\end{table}

\begin{table}[t!]
\centering
\footnotesize
\caption{Per-rule gap rate on release branches, computed as
$\textit{gap} / (\textit{gap} + \textit{already\_fixed})$ over the branches whose
default-branch fix touched that rule.}
\label{tab:gap-rate-by-rule}
\setlength{\tabcolsep}{1mm}
\begin{tabular}{lr@{\hspace{1.6em}}lr}
\toprule
{\bf Rule} & {\bf Rate} & {\bf Rule} & {\bf Rate} \\
\midrule
\texttt{\footnotesize artipacked}            & 93.6\% & \texttt{\footnotesize unsound-contains}        & 53.8\% \\
\texttt{\footnotesize secrets-inherit}       & 92.9\% & \texttt{\footnotesize use-trusted-publishing}  & 53.5\% \\
\texttt{\footnotesize unpinned-uses}         & 91.6\% & \texttt{\footnotesize bot-conditions}          & 45.1\% \\
\texttt{\footnotesize superfluous-actions}   & 84.5\% & \texttt{\footnotesize github-env}              & 36.9\% \\
\texttt{\footnotesize excessive-permissions} & 81.7\% & \texttt{\footnotesize dangerous-triggers}      & 33.0\% \\
\texttt{\footnotesize archived-uses}         & 79.2\% & \texttt{\footnotesize obfuscation}             & 13.5\% \\
\texttt{\footnotesize misfeature}            & 73.2\% & \texttt{\footnotesize insecure-commands}       & 11.8\% \\
\texttt{\footnotesize template-injection}    & 66.4\% & \texttt{\footnotesize unsound-condition}       & 10.3\% \\
\texttt{\footnotesize unpinned-images}       & 62.1\% & \texttt{\footnotesize unredacted-secrets}      &  0.0\% \\
\texttt{\footnotesize overprovisioned-secrets} & 59.6\% &                                &        \\
\bottomrule
\end{tabular}
\end{table}

We apply the security-fix filter to all 1,903,191 commits to obtain the set of default-branch fixes whose security effect is unambiguous. For each such fix we enumerate every release-style branch of the project through the GitHub API, identifying release branches by naming convention (\S\ref{sec:threats}), fetch the workflow at each branch's current \texttt{\small HEAD}, and re-scan it, labeling the branch \textsc{gap} (the fixed weakness is still present), \emph{already-fixed} (absent), or \emph{inapplicable} (the workflow file does not exist), then aggregate to the project level. This is a point-in-time comparison against each branch's current \texttt{\small HEAD}: it measures \emph{whether} a branch is still exposed, while \emph{how} an already-fixed branch got that way (a deliberate backport or an incidental merge) is deferred to RQ2 (\S\ref{sec:study:rq2}).

\smallskip
\noindent\textbf{The gap is pervasive.}
The security-fix filter yields 69,865 default-branch fixes across 24,238 repositories (a mean of 2.9 per repository), each a weakness a maintainer demonstrably knew how to fix. Auditing every one of the 474,490 release branches these fixes could apply to (TABLE~\ref{tab:branches}), 224,392 (47.3\%) still carry the fixed weakness, nearly three times the 78,907 already fixed, while 171,191 (36.1\%) do not contain the workflow file at all. Restricted to the branches that actually carry the workflow, \ie, those on which a backport is even meaningful, the gap rises to 74.0\% (224,392 of 303,299); this is the denominator behind the per-rule rates below. At the project level, 7,172 of 24,238 projects (29.6\%) have at least one release branch the default-branch fix never reached. The gap is not an artifact of a few outliers: it is spread across thousands of projects, and \S\ref{sec:study:rq3} examines how many branches a single fix must reach.

\smallskip
\noindent\textbf{Does the gap persist on active branches?}
The median date of repositories’ last workflow-modifying commits is May 2025, and 70.1\% of repositories modified workflows in 2025. We randomly sample 600 release branches and down-weight older branches by halving their weight for each six-month period since their last commit. Among vulnerable branches, 53.7\% received commits within the past 24 months. The weighted proportion of vulnerable branches is 49.2\%. The rate is similar when we consider only branches updated within the 6, 12, or 24 months before our audit in June 2026 (46.5\%, 47.5\%, and 50.1\%).

\smallskip
\noindent\textbf{Which weaknesses are left behind.}
Normalizing the gap per rule, defined as $\textit{gap}/(\textit{gap}+\textit{already\_fixed})$ over the branches whose default-branch fix touched that rule, exposes which weaknesses fare worst (TABLE~\ref{tab:gap-rate-by-rule}). The supply-chain rules dominate: \texttt{\small artipacked} (93.6\%), \texttt{\small secrets-inherit} (92.9\%), and \texttt{\small unpinned-uses} (91.6\%) are left unpatched on the large majority of branches that once carried them, the very weaknesses behind the \texttt{\small tj-actions} class of attacks (\S\ref{sec:vuln-taxonomy}).

\smallskip
\noindent\textbf{Is the gap real? Validating the oracle.}
The entire gap measurement rests on \textsc{zizmor}'s verdicts, so we validate it along two axes. For \emph{reproducibility}, we pin one version (1.24.1) and scan content-addressed YAML blobs, making every verdict deterministic; \textsc{zizmor}'s rules encode the official hardening guidance of GitHub~\cite{github-actions-hardening}, the OpenSSF Scorecard~\cite{10163720}, and documented CVEs~\cite{cve-2025-30066}, covering the weaknesses of \S\ref{sec:vuln-taxonomy} and further supply-chain and configuration categories. For \emph{accuracy}, two authors independently adjudicated every \textsc{zizmor} finding on, and checked for weaknesses it missed in, a uniform random sample of 200 workflow files: \textsc{zizmor} achieves 96.8\% precision and 99.1\% recall.
In the security-fix filter, a false positive that disappears after a commit makes the commit count as a security fix, even though it removed no real weakness. With this precision, such cases are a small minority.

\finding{The cross-branch backporting gap is pervasive: 47.3\% of audited release branches still carry a weakness already fixed on the default branch (74.0\% of the branches that carry the workflow at all), and 29.6\% of projects have at least one such branch. The supply-chain rules behind recent action-tag attacks are left unpatched on over 90\% of the branches that once carried them.}

\subsection{RQ2: Characterizing Backporting Practice}\label{sec:study:rq2}

For each of the 78,907 already-fixed branches, we walk the workflow file's commit history on that branch to locate the commit where the issue first disappeared, and compare its date with the default-branch fix.
We classify each branch as a \textit{plausible backport} (the branch had the issue and removed it more than a day after the default branch), \textit{same-day} (removed on the same day, most likely an auto-merge),
\textit{independent prior fix} (removed before the default branch), \textit{never had the issue}, or \textit{inconclusive} when the history is too deep to resolve within a bounded per-record budget, \ie, at most 50 commits per file and 16 minutes per check.
For confirmed plausible backports, we compute the lag from the default-branch fix.

\smallskip
\noindent\textbf{Results.} Walking the history of the 78,907 already-fixed branches (TABLE~\ref{tab:refined}), 19,476 (24.7\%) are confirmed plausible backports, distributed across 1,279 projects. A clear majority of the already-fixed branches are same-day removals (41,408, 52.5\%), most likely reflecting an auto-merge from the default branch rather than a deliberate decision to patch the release line. The confirmed backports are slow: the median lag is 243 days, with 1,576 landing within one to three months, 9,065 within three to twelve, and 6,580 taking more than a year, up to a maximum of 2,326 days. Only 1,296 land within a single week, driven by a few security-conscious projects such as \texttt{\small demergent-labs/azle} (466 of its 5,581 confirmed backports). For most release branches, the window of exposure to an already-fixed weakness is measured in months.

\begin{table}[t]
  \centering
  \caption{Refined classification of the 78,907 \textsc{already-fixed}
  branches by file-history analysis.}
  \label{tab:refined}
  \begin{tabular}{lrr}
    \toprule
    \textbf{Refined category} & \textbf{Branches} & \textbf{Share} \\
    \midrule
    \textbf{Plausible backport} (lag $>$ 1 day)         &  \textbf{19,476} & \textbf{24.7\%} \\
    Same-day (likely auto-merge)                   &   41,408 & 52.5\% \\
    Independent prior fix (lag $<-1$ day)          &   5,670 &  7.2\% \\
    Never had the issue                            &    4,020 &  5.1\% \\
    Inconclusive (history too deep)                &    4,582 &  5.8\% \\
    Timed out (budget exceeded)                    &    3,751 &  4.8\% \\
    \bottomrule
  \end{tabular}
\end{table}

\finding{Deliberate backports are rare, concentrated, and slow: 19,476 of
78,907 already-fixed branches (24.7\%) are confirmed plausible backports;
the median lag is 243 days, and although a security-conscious minority lands 1,296 within a single week, the long tail extends past six years.}

\subsection{RQ3: Deriving Backporting Challenges}\label{sec:study:rq3}

\noindent\textbf{The approaches available today.}
Before turning to what makes backporting hard, we measure how far a maintainer would get with the approaches available today, run on a common set of 12,000 unpatched files (TABLE~\ref{tab:baselines}): a verbatim copy-paste of the source diff; a Dependabot-style single-dependency update, which advances one action to a newer version but cannot add the coupled permission or credential edits that travel with a hardening fix; and a general-purpose LLM prompted to rewrite the whole file. These close only 18.0\%, 13.0\%, and 49.5\% (Gemini) or 33.7\% (GPT) of the files, respectively, so each leaves most branches unfixed, and each fails for a different reason. We now name the three dimensions of difficulty the data reveals and tie each to the approach it defeats.

\smallskip\subsubsection{Multiplicity (One Fix Often Spans Many Branches)}

Backporting is challenging because a single default-branch fix may need to be propagated to many release branches. Among the 7,172 projects with at least one unpatched release branch, the mean number of gap branches is 31.3 per repository. Moreover, 752 distinct default-branch fixes are each missing from more than 50 release branches, including \texttt{\small datalens-tech/datalens-ui} (2,971 branches), \texttt{\small datalens-tech/datalens-backend} (555), \texttt{\small demergent-labs/azle} (2,759), \texttt{\small n8n-io/n8n} (434), and \texttt{\small azure/azure-sdk-for-go} (434). These results show that workflow backporting is often a one-to-many problem, where a project may not be fully remediated until the same fix is adapted and applied across tens or even hundreds of release branches.

\smallskip
\noindent\textbf{Implication for tool design.}
Closing this gap by hand would mean opening up to 2,971 pull requests for a single fix and re-resolving conflicts on each, which is why it is often left undone. A tool must instead fan one extracted fix out across all of a project's affected branches autonomously, validating each on its own.

\smallskip\subsubsection{Coupling (One Commit, Multiple Coordinated Edits)}

34.3\% of security fixes (23,965 of 69,865) couple two or more distinct
\textsc{zizmor} rule categories in a single commit, and the
\{\texttt{\small artipacked}, \texttt{\small excessive-permissions}, \texttt{\small unpinned-uses}\}
triple alone accounts for 13.4\% (9,380) of all
security fixes. This shows that real remediations are commit-level atomic units, rather than
single-change patches.

\begin{table}[t]
\centering
\caption{What the approaches available today achieve on the common set of
12{,}000 unpatched files. Pure-LLM counts only patches whose action pins
resolve to real commits.}
\label{tab:baselines}
\begin{tabular}{lrr}
\toprule
{\bf Approach} & {\bf Passed} & {\bf Rate} \\
\midrule
Pure-LLM rewrite (Gemini) & 5{,}940 & 49.5\% \\
Pure-LLM rewrite (GPT)    & 4{,}044 & 33.7\% \\
Copy-paste of source diff & 2{,}158 & 18.0\% \\
Dependabot-style update   & 1{,}565 & 13.0\% \\
\bottomrule
\end{tabular}
\end{table}

\smallskip
\noindent\textbf{Implication for tool design.}
A coupled fix is correct only as a whole, since pinning the action but leaving \texttt{\small persist-credentials} set still leaks the token. A per-dependency updater cannot express this: a Dependabot-style update raises a single action's version but never adds the coupled \texttt{\small permissions} and \texttt{\small persist-credentials} edits, nor moves an already-vetted fix backward onto an older branch, and it closes only 13.0\% of the unpatched files (TABLE~\ref{tab:baselines}). A tool must therefore treat the \emph{commit} as its atomic unit, accepting a backport only if every edit lands, never a partial one.

\smallskip\subsubsection{Structural Divergence (Release Branches Do Not Match the Default Branch in Shape)}

Release-branch YAML structures diverge from default-branch structures
on most projects: job names are renamed, step orderings
differ, and files are refactored. In 60.5\% of the unpatched branches in our dataset, the location the fix occupied on the default branch no longer exists on the release branch, so the fix cannot be applied where it originally sat.
Fig.~\ref{fig:divergence} contrasts a default-branch fix with a
structurally divergent release branch that carries the same weakness.

\begin{figure}[t]
\centering
\begin{minted}[style=colorful, label={Default branch (fix applied)}]{yaml}
jobs:
  build:  # job named "build"
    runs-on: ubuntu-latest
    steps:
      - uses: actions/checkout@b4ff...  # step 0
        with:
          persist-credentials: false
      - run: make build
\end{minted}
\begin{minted}[style=colorful, label={Release branch (same project, different shape)}]{yaml}
jobs:
  prepare:  # different job name
    runs-on: ubuntu-22.04
    steps:
      - run: ./scripts/setup.sh   # extra step
      - uses: actions/checkout@v3 # step 1, unpinned
                                  # no `with` block
      - run: make build
\end{minted}
\caption{Why a textual diff cannot be backported verbatim. The
default-branch patch targets
\texttt{jobs.build.steps[0].with.persist-credentials}, but on the
release branch the job is renamed (\texttt{prepare}), an extra setup step
shifts the checkout to index~1, and no \texttt{with} block exists yet.
}
\label{fig:divergence}
\end{figure}

\smallskip
\noindent\textbf{Implication for tool design.}
Without the fix's literal path, a verbatim copy-paste of the source diff has no pre-image to match and cannot apply; even where a hunk does apply, it reuses the source branch's pinned SHA rather than the version the target runs, which can downgrade or break the branch. Copy-paste closes only 18.0\% of the unpatched files (TABLE~\ref{tab:baselines}). A tool must instead locate the target by structural content rather than position, binding job and step to whatever the branch actually contains, and express each edit as a desired end-state that can create missing structure, rather than as a textual hunk.
Even a general-purpose LLM prompted to rewrite the whole file, which handles the structural rewriting the other two approaches cannot, closes only 49.5\% of the unpatched files (TABLE~\ref{tab:baselines}).
No single existing approach handles multiplicity, coupling, and structural divergence together, and this is the gap \sysname closes.

\finding{Workflow backporting in practice exhibits three operational
dimensions of difficulty: \emph{multiplicity}, \emph{coupling}, and \emph{structural divergence}. The three dimensions are orthogonal:
each must be addressed for a backporting tool to be useful at scale.}

\section{\sysname}
\label{sec:tool}

\sysname takes as input a security fix that has already been applied on the default branch and propagates it to the release branches that still contain the vulnerable workflow. As Fig.~\ref{fig:workflow} shows, it operates in four steps.

\textbf{Step 1 (Compile).} \sysname turns the fixing commit into a \emph{Workflow Semantic Patch} (WSP), which is a program in a small domain-specific language (DSL) that describes the edits the fix makes to the workflow's structure rather than to its text (\S\ref{sec:tool-compile}).

\textbf{Step 2 (Apply).} It replays this single WSP on each release branch that still carries the vulnerability, binding the WSP's anchors to whatever the branch actually contains (\S\ref{sec:tool-apply}).

\textbf{Step 3 (Validate).} It accepts a patch only when four independent oracles agree that the targeted vulnerability is gone and nothing else broke (\S\ref{sec:tool-verify}).

\textbf{Step 4 (LLM fallback).} For the cases the deterministic engine declines or that fail validation, \sysname asks a large language model to synthesize a revised WSP from the engine's counterexamples, holding the result to the same oracles (\S\ref{sec:tool-llm}).

The same WSP that the engine executes is also what a reviewer reads, since it is a short, human-readable program, so every backport is auditable before it becomes a pull request. The rest of this section defines the WSP language and then details each step.

\begin{figure}[t]
\centering
\includegraphics[width=\linewidth]{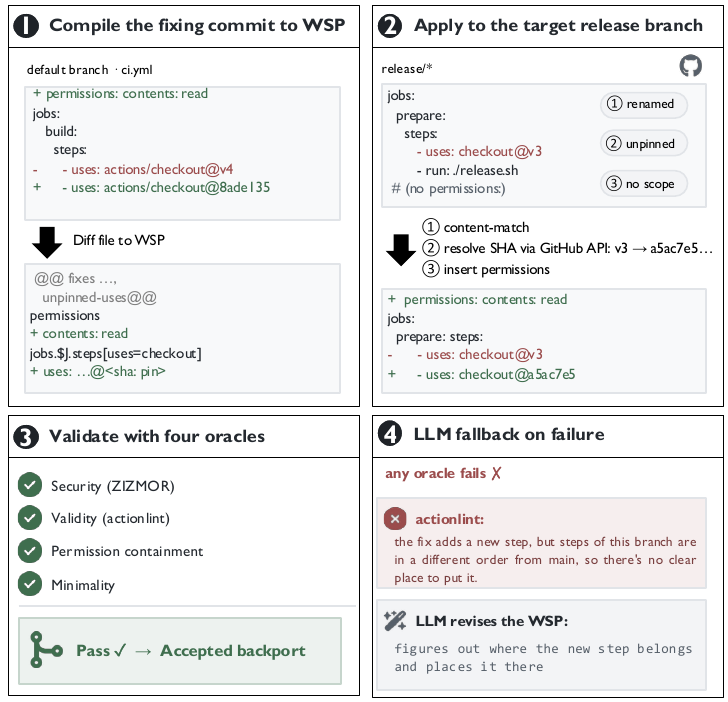}
\caption{Overview of \sysname. A default-branch fix is compiled once into a
Workflow Semantic Patch (WSP), applied to each divergent release branch by
content-based matching, and checked by four independent oracles. Cases the
deterministic engine declines, or that fail validation, are routed to a
counterexample-guided LLM that synthesizes a revised WSP, which is then held to
the same apply-and-verify path.}
\label{fig:workflow}
\end{figure}

\subsection{The Workflow Semantic Patch (WSP)}
\label{sec:tool-wsp}

\begin{figure}[t]
\begin{minted}[fontsize=\scriptsize]{text}
program  ::= '@@' header '@@' block+
header   ::= source_line fixes_line metavar_decl*
fixes_line   ::= 'fixes' rule (',' rule)*
metavar_decl ::= 'metavariable job' var 'pin' string
                  ['recover' fingerprint] 'bind' ('one' | 'each')
block    ::= anchor (edit)+
anchor   ::= seg ('.' seg | '[' field '=' value ']')*
seg      ::= key | var                  // var: a job metavariable
edit     ::= '+' key [':' value]        // ensure_present
           | '-' key                    // ensure_absent
           | '-' key ':' old            // rewrite: a '-' and a '+'
             '+' key ':' new            //   on the same key in one block
value    ::= scalar | json_flow | '<sha: pin target_ref>'
\end{minted}
\begin{minted}[fontsize=\scriptsize]{diff}
@@
# source: <repo>@<sha> .github/workflows/ci.yml
fixes artipacked, excessive-permissions, unpinned-uses
metavariable job $J pin "build" recover uses=actions/checkout bind one
@@

jobs.$J.steps[uses=actions/checkout].with
+ persist-credentials: false

permissions
+ contents: read

jobs.$J.steps[uses=actions/checkout]
- uses: actions/checkout@<tag>
+ uses: actions/checkout@<sha: pin target_ref>
\end{minted}
\caption{The WSP language: its core grammar (top, EBNF, simplified) and the
compiled program that backports a fix (bottom). The header names the master source, lists the \textsc{zizmor} rules
removed, and declares one job metavariable per touched job. Each block is an
\emph{anchor} (a path of mapping keys, a \texttt{\$J} job metavariable, and
content \texttt{[field=value]} selectors) followed by \emph{edits}: a lone
\texttt{+} is \textsc{ensure\_present}, a lone \texttt{-} is
\textsc{ensure\_absent}, a \texttt{-}/\texttt{+} pair on the same key is a
rewrite, and \texttt{<sha:\,pin\,...>} resolves the target branch's own reference
at apply time. The full grammar is in our artifact.}
\label{fig:wsp}
\end{figure}

The WSP is a small domain-specific language (DSL) that records a fix as edits to the workflow's tree structure rather than to its text. Its design is inspired by Coccinelle's Semantic Patch Language (SmPL)~\cite{216013}, a tool that expresses code changes as structural patterns rather than line diffs, from which it borrows the \texttt{\small @@}-delimited header, the \texttt{\small +}/\texttt{\small -} edit lines, and metavariables, i.e., named placeholders that match a piece of structure the patch then edits. It differs in three ways that the workflow-backporting setting demands: a WSP is compiled automatically from one before/after diff rather than hand-written, its edits are idempotent assertions about a desired end state rather than imperative hunks, and it targets YAML and GitHub Actions semantics, namely jobs, steps, and \texttt{\small uses} identity, rather than a C abstract syntax tree.

Fig.~\ref{fig:wsp} gives the core grammar (top) and a compiled example (bottom). A program opens with a header that names the master source, lists the \textsc{zizmor} rules the commit removed, and declares the job metavariables it uses; the body is a sequence of \textit{blocks}, each an \textit{anchor} that points into the workflow tree followed by the \textit{edits} that should hold there.

\smallskip
\noindent\textbf{Anchors.}
An anchor is a path built from three kinds of segment.
A \texttt{\small key} segment descends into a mapping by its literal name, such as \texttt{\small permissions} or \texttt{\small steps}.
A \texttt{\small selector}, written \texttt{\small [uses=actions/checkout]}, picks one element out of a list by matching its content rather than its index, here the \texttt{\small checkout} step inside a job's step list, so the anchor still resolves after the branch has reordered or inserted entries. Steps can also be selected by \texttt{\small [id=...]} or \texttt{\small [name=...]}.
A \texttt{\small job metavariable}, written \texttt{\small \$J} and declared in the header, stands for one job the master fix touched. Its declaration carries its identity: \texttt{\small pin "build"} is the literal job key the master edited, \texttt{\small recover uses=actions/checkout} names a fingerprint step, i.e., a step that no sibling job runs, used to relocate the job if the branch renamed it, since a unique step lets the engine identify the job by what it does rather than by its name, and \texttt{\small bind one} requires it to bind exactly one job, stopping and deferring to human review rather than fanning the edit out to every job. A multi-job fix declares one metavariable per touched job ($\texttt{\small \$J}, \texttt{\small \$J2}, \ldots$), each with its own edits. We anchor at the job level because renamed jobs are the most common form of structural divergence across release branches.

\smallskip
\noindent\textbf{Edits.}
Each edit specifies the desired state of one key, and the operation is derived from the signs in its block. A lone \texttt{\small + key: v} is \textsc{ensure\_present} (the key must be present with value \texttt{\small v}); a lone \texttt{\small - key} is \textsc{ensure\_absent} (the key must be absent); and a \texttt{\small - key: old} paired with a \texttt{\small + key: new} on the same key is a \textsc{rewrite}. Edits are idempotent assertions, not hunks: an assertion that already holds is a no-op, so replaying a WSP onto a partially fixed or drifted branch converges instead of double-applying.
One value form, the reference to an external action, is special. A WSP encodes the security intent behind such a reference rather than the default branch's concrete SHA. The fix on the default branch replaces a mutable tag, e.g., \texttt{\small @v4}, with the exact commit that tag pointed to, which removes the risk that the tag is later rewritten to malicious code. Copying that same commit onto a release branch can be unsafe, however, since the default branch may run a newer action version than the release branch, so the copied commit would silently upgrade the branch. A \texttt{\small uses} pin is therefore written as the marker \texttt{\small <sha: pin target\_ref>}, which tells the engine to read the reference the target branch itself uses and pin it to the exact commit it currently points to. This keeps the branch on its own action version while removing the mutable tag, so the security intent, i.e., no rewritable tag, is preserved without changing which version runs.

\subsection{Master Commit to WSP}
\label{sec:tool-compile}

\sysname compiles the master commit, together with the rules the fix removed, into a WSP. The compiler works structurally, not textually. It first \emph{flattens} the before and after workflows into maps from a structural path to a leaf value, where a path keys each list element by its content identity, for instance \texttt{\small \seqsplit{jobs.build.steps[uses=actions/checkout].with.persist-credentials}}, rather than by its index. Comparing the two maps by set algebra yields the leaves the commit added, removed, and changed. The remaining stages turn these leaves into edits.

\smallskip
\noindent\textbf{Deriving job metavariables.}
The compiler declares one metavariable for each job touched by the diff, using \texttt{\small pin "job" bind one} to anchor it to the literal job key and \texttt{\small recover} to specify a fingerprint: a step whose \texttt{\small uses} action is unique to that job, since a job's name may change on a release branch but the action it runs usually does not. A job with no globally unique step gets no fingerprint and fails closed on rename rather than risk the wrong job. Keeping each touched job on its own metavariable matters when a commit edits several jobs differently, since a shared anchor would fan one job's edit out to the others, widening privileges on jobs the fix never touched.

\smallskip
\noindent\textbf{Consolidating type changes.}
Most leaves become one edit each, but a value that flips between a scalar and a block does not. For example, \texttt{\small secrets: inherit} (a scalar) becoming a \texttt{\small secrets:} mapping with several entries (a block) would decompose into an \textsc{ensure\_absent} on \texttt{\small secrets} and several \textsc{ensure\_present} children, where the \textsc{ensure\_absent} would delete the whole \texttt{\small secrets} key before its children could be written. The compiler detects such scalar-to-block transitions and emits a single rewrite that replaces the old value with the entire new block at once.

\smallskip
\noindent\textbf{Declining whole-step changes.}
An edit can set fields inside an existing step, but it cannot create a step that is not there. When the master fix adds or removes an entire step, the compiler cannot know where that step belongs on a branch whose job has a different set of steps in a different order. Rather than guess, it flags the edit for review and routes the case to the LLM fallback described below; validation later refuses any WSP whose edits did not all land.

\smallskip
\noindent\textbf{Filtering to the security intent.}
A mined security-fix commit is rarely surgical: alongside the security change it often bumps a \texttt{\small run} script, a \texttt{\small with} argument, or a matrix entry, and replaying those onto a release branch would revert content the branch evolved independently. The compiler therefore auto-applies only edits attributable to a \emph{targeted} rule's security construct, through a fixed rule-to-construct map (\eg, \texttt{\small excessive-permissions}\,$\to$\,\texttt{\small permissions}, \texttt{\small unpinned-uses}\,$\to$\,\texttt{\small uses}, \texttt{\small template-injection}\,$\to$\,\{\texttt{\small env},\,\texttt{\small run}\}), where a construct is the specific workflow key a rule concerns. Every other edit is flagged for review and never lands, so a backport carries the fix's security content and nothing else.

\subsection{WSP to Release Branch}
\label{sec:tool-apply}

\sysname applies a WSP by resolving each block's anchor against the release branch's parse tree and then making the block's edits. Resolution walks the anchor segment by segment: a \texttt{\small key} descends by name; the job metavariable binds the \texttt{\small pin}ned job, or, if it was renamed, the one job matching the \texttt{\small recover} fingerprint, with an ambiguous match held for review and the engine never inventing a job; and a \texttt{\small [uses=...]} selector matches a step by content rather than index. Because every segment resolves against the branch's own structure, one compiled WSP applies across branches whose jobs and steps have moved.

A resolved anchor may stop short when only mapping keys are missing, for instance a job with no \texttt{\small permissions} block yet; the engine creates such keys for an \textsc{ensure\_present} edit but never invents a missing job or step. The edits then run in their derived form: \textsc{ensure\_present} sets a key, \textsc{ensure\_absent} deletes one, and \textsc{rewrite} replaces a value. A \texttt{\small uses} pin is resolved here: the engine reads the reference the target step carries, queries GitHub for the commit it points to, and writes that SHA, adding the original tag as a trailing comment for readability; a reference that does not resolve becomes a review item, never a guess. All edits use a format-preserving YAML round-trip, so comments, quoting, and indentation survive.

\subsection{Validation}
\label{sec:tool-verify}

A backport is acceptable only if it removes the targeted weakness without introducing a new one, leaves the workflow runnable, and changes nothing else. \sysname checks this with four oracles that are independent of the WSP engine, and accepts a patch only if all four agree.
\emph{(1) Security.} \sysname re-runs \textsc{zizmor}, the static analyzer used throughout to detect workflow weaknesses, on the target before and after the patch, requiring every edited scope to end free of the targeted findings with no new finding inside it. The check is scoped to the edited regions because a branch may carry other, pre-existing instances of the same rule that the fix never addressed.
\emph{(2) Validity.} \textsc{actionlint}, a linter that checks GitHub Actions workflow syntax and semantics, must report no new finding relative to the target, the strongest static proxy for ``the workflow still runs'', since actually executing the workflow would need the project's secrets and CI environment.
\emph{(3) Permission containment.} No \emph{untouched} job's own \texttt{\small permissions} block may change, catching the over-grant and over-strip that a scanner misses, since any explicit permissions block silences \texttt{\small excessive-permissions} regardless of what it grants.
\emph{(4) Minimality.} A structural diff of the target and patched file must attribute every changed leaf to a targeted rule's construct. This oracle uses no scanner, so it does not simply re-confirm the scanner-based definition of a fix, and it catches a patch that silences the scanner while reverting the branch's own \texttt{\small run} or \texttt{\small with} content.

\subsection{Recovering Declined Cases with LLM Assistance}
\label{sec:tool-llm}

On its own, the deterministic engine cannot handle every fix. It declines a fix that needs a whole step added or removed, because it cannot tell where the step belongs on a branch whose steps differ in number and order, for instance a fix that inserts a hardening step before \texttt{\small checkout} when the target job has a different step sequence. It also declines a fix whose anchor cannot be bound, for instance when the branch renamed the touched job and no longer runs a step that uniquely identifies it, so the job metavariable matches nothing under \texttt{\small bind one}. These are the most divergent cases, where a static rewriter would have to guess. To recover some of them, \sysname turns to an LLM, while holding it to exactly the same oracles it applies to its own patches.

We frame this as counterexample-guided inductive synthesis (CEGIS)~\cite{10.1007/10722167_15,1714168} over the WSP language. The model never edits the workflow; it proposes a WSP, and the engine stays the only component that touches the file, so format preservation, idempotent anchoring, and deterministic pinning still hold. Given the master fix's before and after files, the security-relevant part of its diff, the touched jobs, the target file, and the rules to remove, the model emits a candidate WSP, which \sysname applies and runs through the four oracles. On rejection it returns concrete counterexamples, namely the anchors that did not resolve, the edits it could not place, and each oracle violation, and the model revises. The loop runs for a fixed number of rounds ($K=3$), and a case that never passes is reported as unsolved. Because every candidate is machine-applied and machine-checked, the fallback inherits the same discipline as the engine: a WSP that does not apply and pass the checks is never accepted.

\section{Evaluation}
\label{sec:eval}

We evaluate \sysname on the unpatched workflow files from our study, asking two questions:
\begin{itemize}[leftmargin=*]
  \item \textbf{RQ4 (Comparison).} How does \sysname compare with the approaches a maintainer could use today?
  \item \textbf{RQ5 (Cost).} What are the cost of \sysname compared with general-purpose LLMs?
\end{itemize}

Beyond these two questions, we execute 17 of the accepted backports (\S\ref{sec:exec}) and submit 176 backports as issues to real-world projects (\S\ref{sec:respon}).

\subsection{Setup}
\label{sec:eval-setup}

The evaluation runs on the unpatched workflow files from our study. Each pair of a default-branch security fix and a release branch still carrying the corresponding weakness yields one target file. The 224,392 unpatched release branches yield 312,401 evaluation cases, as multiple default-branch fixes may apply to the same branch. Running every baseline on all 312,401 cases is not affordable, since two baselines are commercial LLMs invoked with a multi-round repair loop that can cost more than 4,000 USD.
We therefore stratify the fixes into three groups by the degree of workflow restructuring, ranging from fixes that only modify existing workflow elements to those that add or remove entire steps, and randomly sample 12,000 cases in proportion to each group’s size.
For each method we report the count and fraction of these 12,000 files for which it produces an accepted backport, accepted only when the four validation oracles agree the targeted weakness is gone and nothing else broke.

\subsection{Implementation Details}

The LLM fallback and the pure-LLM baselines use Gemini-3.1-flash-lite and GPT-5.4-mini (model identifiers \texttt{\small google/gemini-3.1-flash-lite} and \texttt{\small openai/gpt-5.4-mini}), accessed in June 2026. All requests set temperature 0 and a limit of 8,192 output tokens, and leave top-p at its default. In each round, the pure-LLM baselines output a rewrite of the whole file. In contrast, \sysname's fallback generates and revises a WSP, which the symbolic engine applies to the target file. Both approaches allow up to three rounds, with the previous output and validation feedback provided after each unsuccessful attempt. We use \textsc{zizmor} 1.24.1 and \textsc{actionlint} 1.7.12.

\subsection{RQ4: Comparison with Baselines}
\label{sec:eval-rq4}

RQ4 sets \sysname against the three approaches available to a maintainer today: a verbatim copy-paste of the source diff, a Dependabot-style~\cite{dependabot} single-dependency updater, and a general-purpose LLM that rewrites the whole file, whose failure modes RQ3 tied to the three challenges. To keep the LLM comparison fair, the pure-LLM baseline uses the same model as the \sysname configuration it is compared against, so we compare \sysname (Gemini) with pure-LLM (Gemini) and \sysname (GPT) with pure-LLM (GPT); the only difference is whether the model rewrites the whole file or only synthesizes a Workflow Semantic Patch (WSP), which the engine then applies and pins to the branch's live action versions. All methods run on the same 12,000 files, and for the pure-LLM baseline we count a patch only when its action pins resolve to real commits, since an unresolvable pin would not run and so would not help.

\begin{table}[t]
\centering
\caption{RQ4 comparison of \sysname against the approaches available today,
on the common set of 12{,}000 unpatched files. Pure-LLM counts only patches whose
action pins resolve to real commits.}
\label{tab:rq6}
\begin{tabular}{lrr}
\toprule
{\bf Method} & {\bf Passed} & {\bf Rate} \\
\midrule
\sysname (Gemini-3.1-flash-lite) & 8{,}148 & 67.9\% \\
\sysname (GPT-5.4-mini)          & 7{,}174 & 59.8\% \\
Pure-LLM (Gemini)                & 5{,}940 & 49.5\% \\
Pure-LLM (GPT)                   & 4{,}044 & 33.7\% \\
Symbolic engine only             & 3{,}291 & 27.4\% \\
\textsc{zizmor} autofix          & 2{,}664 & 22.2\% \\
Copy-paste                       & 2{,}158 & 18.0\% \\
Dependabot-style update          & 1{,}565 & 13.0\% \\
\bottomrule
\end{tabular}
\end{table}

TABLE~\ref{tab:rq6} reports the result. \sysname accepts 67.9\% of files with
Gemini and 59.8\% with GPT, well above every baseline: 49.5\% (Gemini) and 33.7\% (GPT) for the
same two models used alone, 18.0\% for copy-paste, and 13.0\% for Dependabot. 
Without the LLM fallback, the symbolic engine accepts 27.4\% of the files, already above both naive baselines, copy-paste and Dependabot.
The LLM fallback is invoked only on the 8,709 files the engine declines and adds the remaining 40.5\% (4,857 backports) of the 67.9\% with Gemini. The engine accepts 68.6\% of fixes requiring no workflow restructuring but only 1.1\% of fixes that add or remove entire steps, while the fallback raises this rate to 56.3\%. On these 8,709 files, the fallback accepts 55.8\% with Gemini and 44.6\% with GPT, versus 39.7\% and 18.4\% for the pure-LLM baselines. Because the model only synthesizes a WSP whose pins the engine resolves against the live branch, its patches never fail on a fabricated commit hash.

We also compare with \textsc{zizmor}’s own autofix mode (\texttt{\small zizmor --fix=all}), which closes 22.2\% of the same 12,000 files under the four oracles (TABLE~\ref{tab:rq6}), against 67.9\% for \sysname. In 21.4\% of the files, \textsc{zizmor}’s output removes none of the targeted findings, introduces a new finding, or fails \textsc{actionlint}, and in another 56.4\% it is rejected by the minimality oracle because it introduces unrelated changes. \textsc{zizmor} also has no autofix for \texttt{\small excessive-permissions}, \texttt{\small secrets-inherit}, or \texttt{\small unpinned-images}. Of the 4,338 files whose fix targets at least one of these rules, \sysname's accepted patches remove findings of these rules in 1,743 (40.2\%) with Gemini and 1,583 (36.5\%) with GPT, and the symbolic engine alone does so in 1,061 (24.5\%). This is because \sysname transfers the maintainer’s own remediation, which can capture project-specific requirements. For example, the Gemini patches that fix \texttt{\small excessive-permissions} grant \texttt{\small pull-requests: write} in 288 files and \texttt{\small contents: write} in 193 files. In almost all of these files, the permissions blocks are identical to those in the maintainer’s fix.

On the same 12,000 files, \sysname differs significantly from every baseline under exact McNemar tests with Holm correction ($p < 10^{-50}$). Over the pure-LLM baseline with the same model, \sysname accepts 18.4 percentage points more files with Gemini (95\% CI $[17.1, 19.7]$) and 26.1 more with GPT (95\% CI $[25.0, 27.2]$).

\finding{\sysname accepts up to 67.9\% of unpatched files, about 3.8 times the best naive baseline (copy-paste at 18.0\%) and above \textsc{zizmor}’s autofix (22.2\%) and the pure-LLM baseline (49.5\%) under the same model.}

\begin{table}[t]
\centering
\caption{Per-case LLM cost, averaged over the cases that invoke each path.
Fallback rounds are capped at 3 (GPT mean 2.01, Gemini mean 1.82).}
\label{tab:cost}
\begin{tabular}{lrrr}
\toprule
{\bf Path} & {\bf Cases} & {\bf In tok.} & {\bf Out tok.} \\
\midrule
Fallback WSP (GPT)        &  8{,}709 & 9{,}066 &   446 \\
Fallback WSP (Gemini)     &  8{,}709 & 9{,}032 &   501 \\
Pure-LLM rewrite (GPT)    & 12{,}000 & 3{,}954 & 1{,}158 \\
Pure-LLM rewrite (Gemini) & 12{,}000 & 4{,}551 & 1{,}343 \\
\bottomrule
\end{tabular}
\end{table}

\subsection{RQ5: Cost}
\label{sec:eval-cost}

\sysname concentrates its cost on the cases that need it. The symbolic engine is
deterministic, performing a structural diff and two \textsc{zizmor} scans, before and after the edit, per branch with no LLM API
call, so the files it accepts on its own and the large fan-outs the multiplicity challenge requires, i.e., replaying one fix across a project's many branches
(\eg, 2,971 branches for \texttt{\small datalens-ui}), cost zero tokens and parallelize
freely. The LLM fallback fires only on the 8,709 of 12,000 sampled files that the
engine cannot place, capped at three rounds (mean 2.01 under GPT, 1.82 under
Gemini). Because the model emits a compact WSP (about 450 to 500 output tokens)
rather than a full-file rewrite (1,200 to 1,340), and output tokens dominate API
cost (TABLE~\ref{tab:cost}), the entire evaluation runs at well under one US cent
per accepted backport.

\finding{The deterministic engine accepts 27.4\% of the files at zero LLM API cost,
including fan-outs of thousands of branches, and the fallback runs only on the files the engine cannot place, at a few hundred output tokens per case, keeping the per-backport cost well below one cent.}

\subsection{Execution Checks}\label{sec:exec}

We cannot execute all patched workflows because doing so would require access to each project's secrets, runners, and triggering events. Due to limited time, we randomly selected and executed 17 workflows that could be completed within the available time, and all 17 \sysname-patched versions completed successfully. By design, the symbolic engine does not generate new code; it transfers maintainer-authored edits from the default branch, reducing the risk of unintended behavior.

\subsection{Maintainer Responses}\label{sec:respon}

We selected non-archived projects, prioritizing those whose vulnerable release branches were most recently updated, and submitted 176 issues, each including \sysname's generated patch. Of these, 44 have been confirmed; 16 have been remediated, including 2 merged patches and 14 affected branches removed, with another 8 fixes under review. The merged backports, covering 6 workflow files, provide real-world evidence of patch correctness, as they match \sysname's patches and execute successfully in the maintainers’ environments. Among the remaining 132 submissions, 62 were declined because maintainers had no plans to maintain the branch (34), considered the reported weakness not applicable to their use case or branch (22), or did not accept reports generated by automated tools (6). The remaining 70 have not yet received a response.
Because most remediations removed the affected branch and the largest group of declined issues concerns branches the maintainers had no plans to maintain, \sysname is most useful on the release branches that a project still maintains.

\section{Threats to Validity}
\label{sec:threats}

\noindent\textbf{Internal validity.} Both the study and \sysname rely on \textsc{zizmor}, a static analyzer for GitHub Actions, to identify security fixes and to decide whether a backport removes the targeted weakness, so our results inherit its accuracy. Manual validation on a random sample of 200 workflow files showed that \textsc{zizmor} achieves 96.8\% precision and 99.1\% recall. We applied the same validation to \textsc{actionlint}, the linter \sysname uses to confirm a patched workflow remains syntactically valid, obtaining 97.9\% precision and 99.4\% recall.

\vspace{0.1cm}
\noindent\textbf{External validity.} Our corpus comes from public GitHub Actions histories and skews toward active, well-maintained projects, so behavior on dormant repositories may differ. We identify release branches from naming conventions (e.g., \texttt{\small release/*} or version-tagged names), so projects that name their maintenance branches differently are missed and their gaps go uncounted.
What's more, our evaluation runs on a sample of 12,000 of the 312,401 cases (\S\ref{sec:eval-setup}). To assess representativeness, we further ran \sysname with the lower-cost DeepSeek-V4-Flash on all 312,401 cases. It accepts 66.5\% of the full population versus 67.3\% on the sample, indicating that the sample closely reflects the full population.

\vspace{0.1cm}
\noindent\textbf{Construct validity.} We count a release branch as backported only when the fix lands more than a day after the default-branch fix, excluding same-day removals, which we cannot distinguish from automatic merges, so our backport counts are a lower bound. Acceptance is a static proxy for a correct and runnable patch: \textsc{zizmor} confirms the targeted weakness is gone with nothing new flagged, and \textsc{actionlint} checks that the patched workflow remains syntactically valid, standing in for execution.
We executed 17 randomly selected accepted backports that could be completed within the available time, and all 17 completed successfully (\S\ref{sec:exec}); the merged backports, covering 6 workflow files, also execute successfully in the maintainers’ environments (\S\ref{sec:respon}).
We do not run the other patched workflows themselves, which would require each project's secrets and CI/CD infrastructure, so a patch that passes our oracles may still fail at runtime.
\section{Related Work}
\label{sec:related}

\noindent\textbf{Empirical studies of GitHub Actions.} Prior work characterizes Actions adoption and operation: Decan \etal~\cite{9978190} find 43.9\% of nearly 70K repositories use Actions, Golzadeh \etal~\cite{9825792} link its rise to the decline of Travis, CircleCI, and Azure, and Kinsman \etal~\cite{9463074} study activity shifts after adoption. Saroar and Nayebi~\cite{10.1145/3593434.3593475} survey practitioner perceptions, Valenzuela-Toledo and Bergel~\cite{9825774} study workflow maintenance, and Zheng \etal~\cite{10.1145/3749371} taxonomize 375 failed runs across 260 projects. None of these ask whether a default-branch fix reaches the release branches downstream users build from, which is the gap our study quantifies.

\vspace{0.1cm}
\noindent\textbf{Workflow and CI/CD security.} Static analyzers such as \textsc{zizmor}~\cite{zizmor}, GHAST~\cite{10.1145/3560835.3564554}, ARGUS~\cite{291265}, and those surveyed by Fares \etal~\cite{fares2026unpackingsecurityscannersgithub} detect weaknesses within a single, isolated workflow file. Large-scale audits confirm such weaknesses are pervasive~\cite{281396,10.1145/3560835.3564554,10.1145/3643991.3644899,10061526}, sitting within a broader supply-chain risk mapped by studies of malicious packages~\cite{10.1007/978-3-030-52683-2_2}, the npm threat surface~\cite{236368}, and attack taxonomies~\cite{10179304}, and made concrete by incidents such as CVE-2025-30066~\cite{cve-2025-30066}. This line of work reports that a branch is vulnerable but does not transfer a fix across branches, which is the gap \sysname closes.

\vspace{0.1cm}
\noindent\textbf{Dependency updating.} Automated updaters such as Dependabot~\cite{dependabot} and Renovate~\cite{renovate} raise one pull request per dependency, moving forward toward the latest release. Empirical work shows the limits of this model: vulnerabilities propagate widely through the npm dependency network~\cite{10.1145/3196398.3196401}, fixes reach downstream consumers only after long lags~\cite{10.1007/s10664-021-09951-x}, an ecosystem-level analogue of the 243-day median backport lag we observe, and automated pull requests often overwhelm developers~\cite{8115621}. This single-dependency, forward-only model cannot express a coupled hardening fix, nor move a vetted fix back onto a divergent release branch, and it closes the smallest share of the gap among the methods we evaluate.

\vspace{0.1cm}
\noindent\textbf{Patch backporting and transplantation.} Transferring a fix across versions or programs is a decade-old line of work for source code: LASE applies a learned edit across many locations~\cite{6606596}, muScalpel and CodeCarbonCopy transplant functionality between programs~\cite{10.1007/978-3-319-22183-0_21,10.1145/3106237.3106269}, PatchWeave transplants a patch into a semantically similar program~\cite{10.1145/3412376}, and FixMorph and TSBPORT backport Linux kernel patches~\cite{10.1145/3460319.3464821,10.1145/3576915.3623188}. Studies of backporting practice confirm the task is hard and error-prone~\cite{7816469,7371961,6693095,10173889}. These techniques rely on programming-language semantics, ASTs, types, control- and data-flow, and a compiler or test suite, none of which a YAML workflow provides. Nor do they face the problem workflow backporting requires: resolving external state by pinning an action to the commit its tag currently points to.
Meanwhile, related work evaluates SZZ implementations for tracing bug-fixing commits to bug-inducing commits~\cite{10541859}, while we examine how fixes propagate across release branches.

\vspace{0.1cm}
\noindent\textbf{LLM-based repair.} Where YAML semantics are thin, LLM-based repair, such as zero-shot LLMs that rewrite vulnerable code~\cite{10179324,10172803} or LLM-based agents that drive automated program repair~\cite{10.1145/3650212.3680384,11029735}, remains relevant, and Hou \etal~\cite{10.1145/3695988} survey their broad programming capability.
Empirical studies further examine how practitioners build and evaluate SE agents~\cite{lyu2026practitionersbuildseagents} and what users expect from AI coding assistants~\cite{11334533}.
However, LLMs cannot recover mutable external state, such as an action's live commit SHA, and instead hallucinate it. \sysname uses a model only to synthesize a WSP and leaves SHA resolution to the deterministic engine, so no accepted patch rests on a fabricated SHA.
\section{Conclusion}
\label{sec:conclusion}

We presented the first large-scale study of security-fix backporting in GitHub
Actions workflows. Across 1,903,191 commits we identified 69,865 security fixes in
24,238 repositories, and found the resulting backporting gap pervasive and slow: 47.3\% of
audited release branches remain vulnerable, 29.6\% of projects are affected, and
where a fix is carried over, the median lag is 243 days, with deliberate backports
rare and concentrated in a small number of security-conscious projects. We therefore built \sysname, a neuro-symbolic
backporting tool: a deterministic engine compiles a fix once into a Workflow
Semantic Patch, a structural representation of the edits, and applies it across branches under scanner validation, while an
LLM fallback, held to the same scanner validation and never trusted to edit the file directly, recovers the structurally hardest
cases. \sysname produces a scanner-checked backport for up to 67.9\% of
unpatched files, about 3.8$\times$ the best naive baseline and well above a pure LLM rewrite.
We release the dataset and tool to support further work on
workflow-level supply-chain remediation. Looking ahead, we plan to adapt \sysname to other
configuration-as-code formats such as GitLab CI and strengthen the validity signal beyond what
scanners and linters provide. 

\section*{Data Availability}

The replication package is available at: \url{https://github.com/CharlieMCY/WorkflowFix}.

\section*{Acknowledgments}

This research is supported by the National Research Foundation, Singapore, and the Cyber Security Agency of Singapore under its National Cybersecurity R\&D Programme (Proposal ID: NCR25-DeSCEmT-SMU). Any opinions, findings and conclusions or recommendations expressed in this material are those of the author(s) and do not reflect the views of the National Research Foundation, Singapore, and the Cyber Security Agency of Singapore.

\bibliographystyle{IEEEtran}
\bibliography{bibtex/references}
\balance

\end{document}